\documentclass{webofc}

\usepackage[varg]{txfonts}   
\usepackage{hyperref}
\usepackage{lineno}
\usepackage{url}

\hypersetup{colorlinks=true,citecolor=blue,urlcolor=blue,linkcolor=blue}
\begin{document}
%
\title{Open Data at ATLAS: Bringing TeV collisions to the World}

\author{\firstname{Giovanni} \lastname{Guerrieri}\inst{1}\fnsep\thanks{\email{giovanni.guerrieri@cern.ch}}, on behalf of the ATLAS Collaboration
}

\institute{CERN, European Laboratory for Particle Physics, Geneva, Switzerland
          }

\abstract{ATLAS Open Data for Education delivers proton--proton collision data from the ATLAS experiment at CERN to the public along with open-access resources for education and outreach. To date ATLAS has released a substantial amount of data from 8 TeV and 13 TeV collisions in an easily-accessible format and supported by dedicated documentation, software, and tutorials to ensure that everyone can access and exploit the data for different educational objectives. Along with datasets, ATLAS also provides data visualisation tools and interactive web based applications for studying the data, along with Jupyter Notebooks and downloadable code enabling users to further analyse data for known and unknown physics cases. The Open Data educational platform which hosts the data and tools is used by tens of thousands of students worldwide, and we present the project development, lessons learnt, impacts, and future goals.
}
\maketitle
{\let\thefootnote\relax\footnote{\hspace{-5.8mm} Copyright 2025 CERN for the benefit of the ATLAS Collaboration. \\ Reproduction of this article or parts of it is
allowed as specified in the CC-BY-4.0 license.}}

\section{Introduction}
\label{intro}
Open data constitutes one of the most effective means for scientific collaborations to disseminate knowledge and establish a long lasting legacy.
It serves a dual purpose, not only enabling and advancing research outside of the lifetime of the experiments, but also fostering education by providing students and educators with authentic datasets to explore and learn from.

The ATLAS Open Data initiative~\cite{AOD} covers both education and research efforts. 
The education project, which is the focus of these proceedings, offers open access to proton--proton collision data gathered by the ATLAS experiment~\cite{atlas} at the Large Hadron Collider (LHC)~\cite{LHC}. The provided resources are developed collaboratively with students and educators, and aim at high school, undergraduate, and postgraduate learners, as well as individuals who are not formally associated with any institution. 
On the other hand, the recent Open Data for research release~\cite{research} comes with sufficient detail to be used for new scientific publications.

The project implements the FAIR principles (Findable, Accessible, Interoperable, Reusable)~\cite{fair}, by providing unique identifiers to the resources, adopting standardised protocols for data management, and supporting cross-platform interoperability. These practices integrate well with ATLAS Open Data's core values of accessibility, usability, and transferable expertise. The resources are designed to be inclusive, enabling a wide range of users, regardless of technical background or location, to achieve diverse learning objectives. Beyond High Energy Physics (HEP), participants gain valuable skills in software development and machine learning. The open datasets and related tools are widely utilised in classrooms, public events, masterclasses, and international workshops.
An overview of the past and future Open Data releases for education is presented in Section~\ref{sec-1}. 
Section~\ref{sec-software} describes the tools and the software available for users, whereas the infrastructure and platforms on which the software operates are covered in Section~\ref{sec-infrastructure}.
In Section~\ref{sec-best-practices} the best practices to ensure usability in Open Data are discussed.
Section~\ref{sec-conclusion} summarises the overall ATLAS Open Data activity and describes the future plans.

\section{The ATLAS Open Data releases for Education}
\label{sec-1}
The ATLAS Open Data releases feature a collection of datasets gathered during the ATLAS detector’s data acquisition runs\addtocounter{footnote}{-1}\footnote{The term ``Run'' is used to describe a specific period of data taking, which can last for several years, in which both the experiments and the LHC are operated under reasonably stable conditions.} at the LHC. These datasets include 8 TeV and 13 TeV proton--proton collision data, with both real collision events and Monte Carlo simulations, along with dataset variations to estimate systematic uncertainties.

\begin{figure}[h]
\centering
\includegraphics[width=7.8cm,clip]{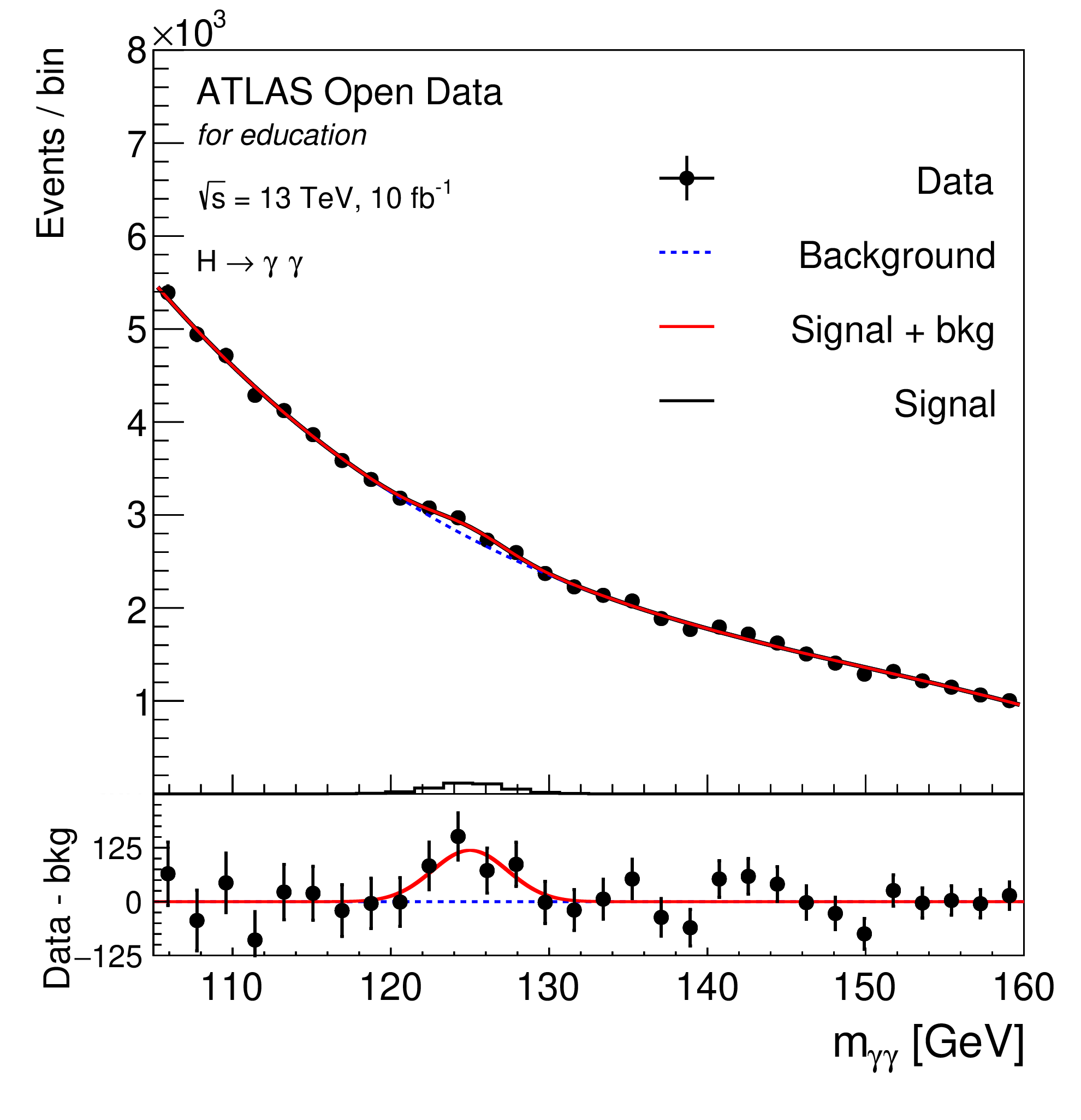}
\caption{Example of analysis performed with the 2020 release for education. Figure found in Ref.~\cite{13tev}.}
\label{fig-myy}
\end{figure}

The datasets are curated with calibrated and simplified information about reconstructed physics objects, making them accessible and manageable for a wide range of users. The inclusion of the “\textit{for education}” label in plot titles such as the one shown in Fig.~\ref{fig-myy} ensures clarity of purpose, distinguishing these datasets as tools tailored for learning and exploration.

The releases include:  
\begin{itemize}
    \item \textbf{8 TeV (2016)}~\cite{8tev,8tev-1}: this release features 1 fb$^{-1}$ of LHC Run 1 data, representing approximately 4.5\% of the 2012 dataset ($\sim$6 GB).
    \item \textbf{13 TeV (2020)}~\cite{13tev}: datasets released as part of the LHC Run 2 data-taking, featuring 10~fb$^{-1}$ of data, equivalent to about 30\% of the 2016 dataset ($\sim$150 GB).
    \item \textbf{13 TeV (2025)}: The upcoming release will offer 36 fb$^{-1}$ of data, further enhancing the scope for educational analysis. 
    The data will be provided as a flat ROOT \texttt{NTuple}~\cite{root}.
\end{itemize}

\subsection{Resources location and preservation}
\label{sec-location}
Each release is accompanied by datasets hosted on the CERN Open Data Portal~\cite{COP} and documentation, tutorials, and supporting resources available on the ATLAS Open Data website~\cite{AOD}. These materials guide learners in analysing particle-physics data in educational environments, offering a hands-on experience that emulates real-world research processes.

All data are shared under a Creative Commons CC0 waiver~\cite{CC0}, with unique DOI identifiers assigned to facilitate proper citation in scientific works. The ATLAS-approved publications mentioned above detail the content, properties, and recommended usage of the datasets. While simplified for educational purposes, the datasets retain complexity and require an undergraduate-level understanding of particle physics.

\newpage
\section{Software and Tools}
\label{sec-software}
To complement the open datasets, the ATLAS collaboration provides a suite of software and tools~\cite{outreach-github} to enable users to analyse and interpret HEP data. Among these tools is the Histogram Analyser, a web-based platform designed to quickly and intuitively interact with cut-based data analysis. This tool allows users to visualise datasets through interactive histograms, focusing on how physicists distinguish between different physics processes. By adjusting the range of data for different variables, users can isolate specific signals, such as Higgs boson events, from background processes. This hands-on approach is particularly useful for building expertise in interpreting data distributions and optimising selection criteria for advanced analyses.

Accompanying the Histogram Analyser is a comprehensive collection of Jupyter notebooks~\cite{jupyter} designed for interactive data analysis. These notebooks cover a wide range of topics, from Standard Model (SM) analyses to Beyond the Standard Model (BSM) physics.

The notebooks are meant to be accessible directly through a web browser, and target users with with varying levels of expertise. They also support multiple programming frameworks, including Python, C++, RDataFrame~\cite{rdf}, and Uproot/Coffea~\cite{uproot, coffea}, ensuring that users can work in a flexible environment and have examples in the environment with which they are most familiar. Beyond physics applications, these resources include tools for training and statistical analyses that extend to other fields, such as data science and related curricula. This versatility makes the ATLAS Open Data resources valuable for broader educational purposes.

A series of YouTube tutorials~\cite{yt} is also provided, offering step-by-step guidance on using the tools, conducting analyses, and understanding the datasets. 

\section{Computing infrastructure}
\label{sec-infrastructure}
Extensive support for computational resources and infrastructure is provided, to accommodate a variety of use cases and user expertise levels, offering web-based (online) resources, and hybrid solutions for limited internet access, that also allow fully offline functionality. The main components of the infrastructure are summarised below, highlighting their capabilities and suitability for different scenarios.

\subsection{Web-based interactive analysis}

\textbf{CERN's SWAN} (\textit{Service for Web-based Analysis})~\cite{swan}, the \textbf{ESCAPE's VRE} (\textit{Virtual Research Environment})~\cite{vre}, and \textbf{Binder}~\cite{binder}, are three platforms offering web-based environments for interactive data analysis. They all leverage Jupyter notebook technologies, enabling users to write and execute code while visualising results inline.

\textbf{SWAN} is a CERN-developed platform that combines advanced computing infrastructure with seamless integration into CERN services, such as CERNBox~\cite{cernbox} for data storage. Supporting Python, C++, and ROOT, SWAN provides the tools necessary for analyses. Its integration with CERN's infrastructure ensures access to robust computing capabilities and powerful data management resources, such as Rucio~\cite{rucio}. However, SWAN requires CERN credentials, which limits its accessibility to external users.

The \textbf{Virtual Research Environment (VRE)} addresses this limitation by providing a platform similar to SWAN but accessible to non-CERN users. It enables users to create accounts without CERN credentials, by using the INDIGO IAM token issuer. By bridging the accessibility gap, the VRE extends the reach of web-based data analysis to a broader audience while maintaining the relevant computational capabilities.

In contrast, \textbf{Binder} provides a lightweight alternative independent of authentication services, making it globally accessible without the need for credentials. Users supply a GitHub repository containing their notebooks, and Binder generates a Docker image to execute the analysis, all at the click of a button.
Its accessible nature and reusable links foster collaboration, though its computational resources are more limited than those of SWAN or the VRE.


\subsection{Docker}
Docker resources~\cite{docker} enhance reproducibility and ease of setup for ATLAS Open Data analyses. The project provides pre-configured Docker images, distributed via the GitHub container registry associated with the ATLAS Open Data repository. Docker’s containerisation ensures that analyses can be run consistently across different systems, resolving common compatibility issues.

Docker is particularly advantageous for collaborative and educational settings, where maintaining consistency and minimising setup complexity are crucial. By shipping all dependencies and settings in a container, users can easily share analytical environments locally or on analysis facilities such as SWAN.

\subsection{Virtual Machines}
The ATLAS Open Data project provides \textbf{Virtual Machines (VMs)} as an offline option for analysing datasets. These VMs replicate a virtual operating system on the user’s machine, preloaded with the necessary 13 TeV software and tools. They offer a viable, low-effort solution for offline analysis, specifically for educational use or environments with unreliable internet access.

\subsection{Main Features Across Platforms}
By offering a diverse set of tools, the ATLAS Open Data infrastructure ensures that users have access to the resources they need to perform analyses, whether online, offline, or in collaborative environments. The main advantages of these resources include:
\begin{itemize}
    \item \textbf{Accessibility:} Tools like Binder prioritise global accessibility, while SWAN and Docker offer high-performance options for users with specific credentials or advanced needs.
    \item \textbf{Reproducibility:} Docker and VMs ensure modular, consistent environments for robust and reproducible analyses.
    \item \textbf{Flexibility:} The listed platforms support various coding languages, as well as accommodating different environments and analysis setups.
    \item \textbf{Collaboration:} Analysis facilities such as SWAN or the VRE enable collaborative work by supporting the sharing of data, software, and computational environments among users.
\end{itemize}

\section{Ensuring the Usability of Open Data: Key Practices and Challenges}
\label{sec-best-practices}
Several factors can significantly impact the usability of Open Data initiatives. Data may become inaccessible if its location changes or if it gets corrupted without proper tracking or backup systems in place. Outdated or deprecated analysis tools and dependencies can potentially limit the effectiveness of the tools. Insufficient or outdated documentation can lead to confusion, making it harder for users to navigate the data and tools. Access restrictions, due to permission settings or infrastructure limitations, can also create barriers, preventing users from using Open Data resources. Furthermore, without sufficient maintenance, infrastructure can become prone to errors, inconsistencies, and prolonged downtime, which disrupts user workflows and data accessibility.

To mitigate these challenges and ensure long-term usability, several best practices should be implemented and maintained throughout the lifecycle of Open Data initiatives:

\begin{itemize}
    \item \textbf{Manage Code in Versioned Repositories}: 
    storing code in version-controlled repositories, such as Git, ensures that changes are tracked and previous versions can be accessed. This practice allows users to work with reliable, reproducible and consistent code.

    \item \textbf{Package the Analysis Environment in Software Containers}:
    containers, such as Docker, encapsulate the analysis environment, including dependencies, tools, and code, ensuring that the environment is reproducible across different systems. This minimises configuration inconsistencies and guarantees that analyses can be performed consistently, regardless of the underlying infrastructure.

    \item \textbf{Document Everything from the Start}:
    comprehensive and up-to-date documentation is essential for usability, but it goes beyond providing instructions. Integrating documentation with issue tracking allows issues to be systematically recorded and addressed. This approach helps maintain clarity and alignment between tools, methods, workflows, and user feedback.

    \item \textbf{Define Easily Reusable Workflows}:
    workflows should be modular and designed with reuse in mind. While notebooks simplify this process, it remains essential to adopt a structured approach that emphasizes consistency. Standardising the organisation, documentation, and execution of workflows ensures they are straightforward to adapt and share across different projects or teams.

    \item \textbf{Use Continuous, Automated Testing}:
    implementing continuous integration and automated testing ensures that updates to the codebase do not introduce errors or break existing functionality. Automated tests help verify the correctness of analysis workflows, and enable timely identification of problems before they impact users.
\end{itemize}

By following these best practices, Open Data efforts can significantly reduce the likelihood of issues that hinder usability, ensuring a more efficient, user-friendly environment. However, it is recommended to also remain committed to vigilance, which is a crucial requirement in order to tackle unforeseen challenges.

\section{Conclusions}
\label{sec-conclusion}
Open Data initiatives in ATLAS have been serving educational purposes since 2016, providing computing resources, tutorials, and documentation that support the datasets. These assets are widely used by several institutions for training sessions, workshops, and masterclasses, fostering the growth of HEP-based learning across disciplines.

A new 13 TeV Open Data for education release is coming soon, offering opportunities to adopt new data formats and expand the available dataset to higher integrated luminosities.
There are also new possibilities emerging from the recent Open Data for research release. It is crucial to identify synergies and complementarities between the education and research efforts. This not only includes adopting shared practices in providing documentation and resources, but also planning for events such as workshops and hackathons; a common ATLAS education + research Open Data event is currently in the organisation phase.

Monitoring and evaluation efforts are crucial to secure a bottom-up, community-based approach towards activities.
Alongside usage statistics and user surveys, an ongoing priority is to keep improving the documentation. Data alone has a limited scope; an equal value resides in the users' ability to access, understand, and use resources effectively.

Finally, the Open Data community is expanding, with new examples and contributions being collected from around the world. This collaborative approach ensures that the ATLAS Open Data initiative continues to evolve and serve a wide range of needs and use cases.

%
%
%

\end{document}